\documentclass[lettersize,journal]{IEEEtran}

\usepackage[utf8]{inputenc}
\usepackage[numbers]{natbib}
\usepackage{wrapfig}

\usepackage{algorithm}
\usepackage[noend]{algpseudocode}
\usepackage{graphicx}
\usepackage{stfloats}
\usepackage{amsfonts}
\usepackage{amsmath}
\usepackage{bm}
\usepackage{cleveref}
\usepackage{color}
\usepackage{multicol}
\usepackage{multirow}
\usepackage{booktabs}
\usepackage{balance} 

\usepackage[utf8]{inputenc}

\def\eg{\emph{e.g.}} 
\def\ie{\emph{i.e.}}

\begin{document}

\title{On the Utility of External Agent Intention Predictor for Human-AI Coordination}

\author {
    Chenxu Wang, Zilong Chen, Angelo Cangelosi and Huaping Liu$^{\ast}$
    \thanks{*Corresponding author}
    \thanks{Chenxu Wang, Zilong Chen and Huaping Liu are with Department of Computer Science and Technology, Tsinghua University, Beijing, China. Angelo Cangelosi is with the University of Manchester, U.K.}
}

\maketitle

\begin{abstract}
Reaching a consensus on the team plans is vital to human-AI coordination. Although previous studies provide approaches through communications in various ways, it could still be hard to coordinate when the AI has no explainable plan to communicate. 
To cover this gap, we suggest incorporating external models to assist humans in understanding the intentions of AI agents. In this paper, we propose a two-stage paradigm that first trains a Theory of Mind (ToM) model from collected offline trajectories of the target agent, and utilizes the model in the process of human-AI collaboration by real-timely displaying the future action predictions of the target agent. Such a paradigm leaves the AI agent as a black box and thus is available for improving any agents. To test our paradigm, we further implement a transformer-based predictor as the ToM model and develop an extended online human-AI collaboration platform for experiments. The comprehensive experimental results verify that human-AI teams can achieve better performance with the help of our model. A user assessment attached to the experiment further demonstrates that our paradigm can significantly enhance the situational awareness of humans. Our study presents the potential to augment the ability of humans via external assistance in human-AI collaboration, which may further inspire future research.

\end{abstract}


\section{Introduction}
\begin{figure*}[t]
\centering
\includegraphics[width=0.9\textwidth]{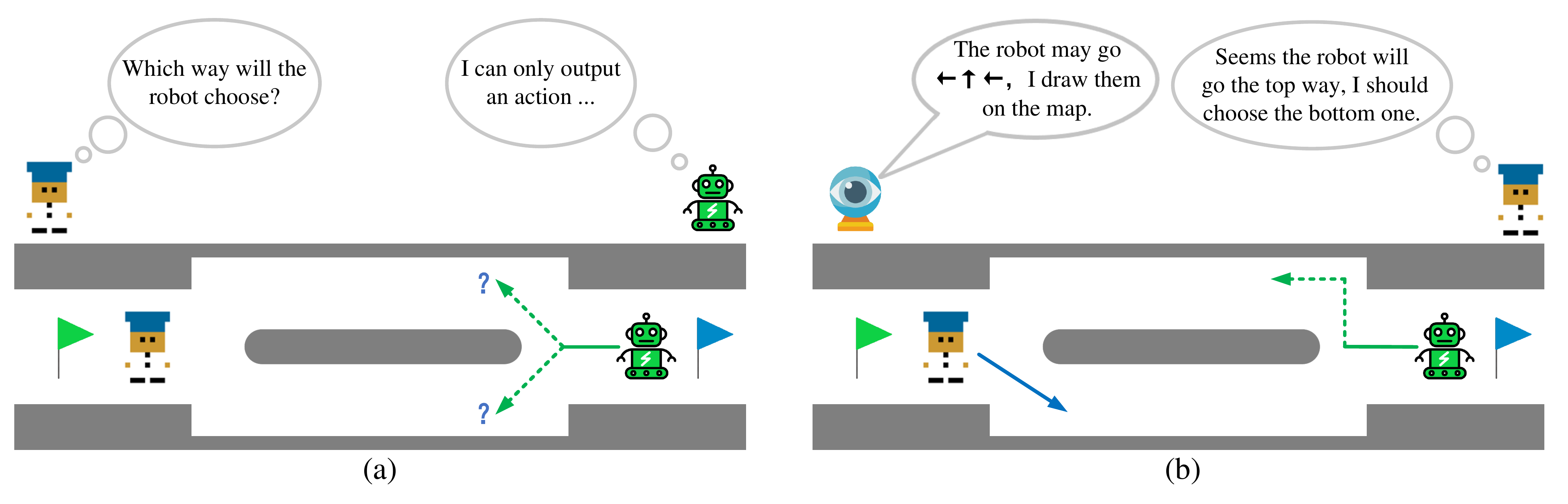}
\caption{A representative case in which the human and the robot want to switch their positions. The flags are the destinations of the corresponding character with respect to the colors, and the gray area denotes the wall or obstacle. The lines denote the human's guess of the routes, whereas the full lines stand for higher confidence. The potential difficulty in the collaboration is reaching an agreement on the selection of routes without collisions.}
\label{fig: example}
\end{figure*}

With the flourishing growth of AI, Human-AI Collaboration has been receiving increasing research interest \cite{dellermann2019future, sowa2021cobots, bcp} and the relevant techniques are applied in various domains, such as robotics \cite{murata2018learning, ghadirzadeh2020human, ibarz2021train}, data science \cite{wang2019human}, and decision making \cite{malhi2020explainable, ibrahim2023explanations}.
Unlike competitive or solo tasks, AI agents and humans share the same goal in collaboration tasks. This naturally encourages each individual to think over the plan and avoid conflict to coordinate better with their teammates to get a higher payoff \cite{lyu2022efficient, aldini2022detection}. Therefore, building a shared mental model to understand the intention of partners is fundamentally essential, which enables all teammates to reach a consensus on task allocations and route planning, and may be vital to collaboration. Consider a simple case shown in \Cref{fig: example}(a), in which a human and a robot need to interchange their positions in a narrow place. To pass the in-between obstacle, they must reach a consensus on the route selection to avoid collisions. However, this could be difficult for the team since they have no access to the other's plan. 

An intuitive way to resolve this dilemma is communication, which is natural in human-human collaboration and has been widely applied in Human-Agent Collaboration \cite{ajoudani2018progress, tabrez2022descriptive, porteous2023communicating}. Typically, the team may communicate through graphical or textual ways to align their mental states \cite{yuan2022situ}. However, as the tasks and environments become complex and erratic, traditional planning-based models are not always feasible. When it comes to Deep Reinforcement Learning (DRL), a class of algorithms that have been widely used in many challenging robotic tasks \cite{ibarz2021train, nguyen2019review, chen2017decentralized}, the model may be opaque and only output an atomic action at each time step with no human-explainable plans to show. Though the explainable AI has been proven important for human-AI collaboration \cite{paleja2021utility}, there may exist a trade-off between the explainability and performance \cite{edmonds2019tale, gunning2019xai}. Prior works also substantiate the importance of communicating about future actions in human-robot collaboration \cite{maccio2022mixed, rosen2020communicating}, however, those actions are usually determined in advance and are not applicable in real-time interactive collaboration, let alone deep learning models that generate atom actions. Therefore, how to conduct communication when the AI agent is not explainable is still an open problem. 


To mitigate this predicament, we suggest building an external Theory of Mind (ToM) model to predict the future actions of the AI agent and assist humans in recognizing its intentions. As the example shown in \Cref{fig: example}(b), the ToM model which has watched past records of the robot can predict the future action of the robot (\textit{left, up, left}), implying the robot will choose the top way. The predictions can then be presented in appropriate ways, \eg, drawn on the map, to help the human better coordinate with the robot. 

\begin{figure}[t]
\centering
\includegraphics[width=0.4\textwidth]{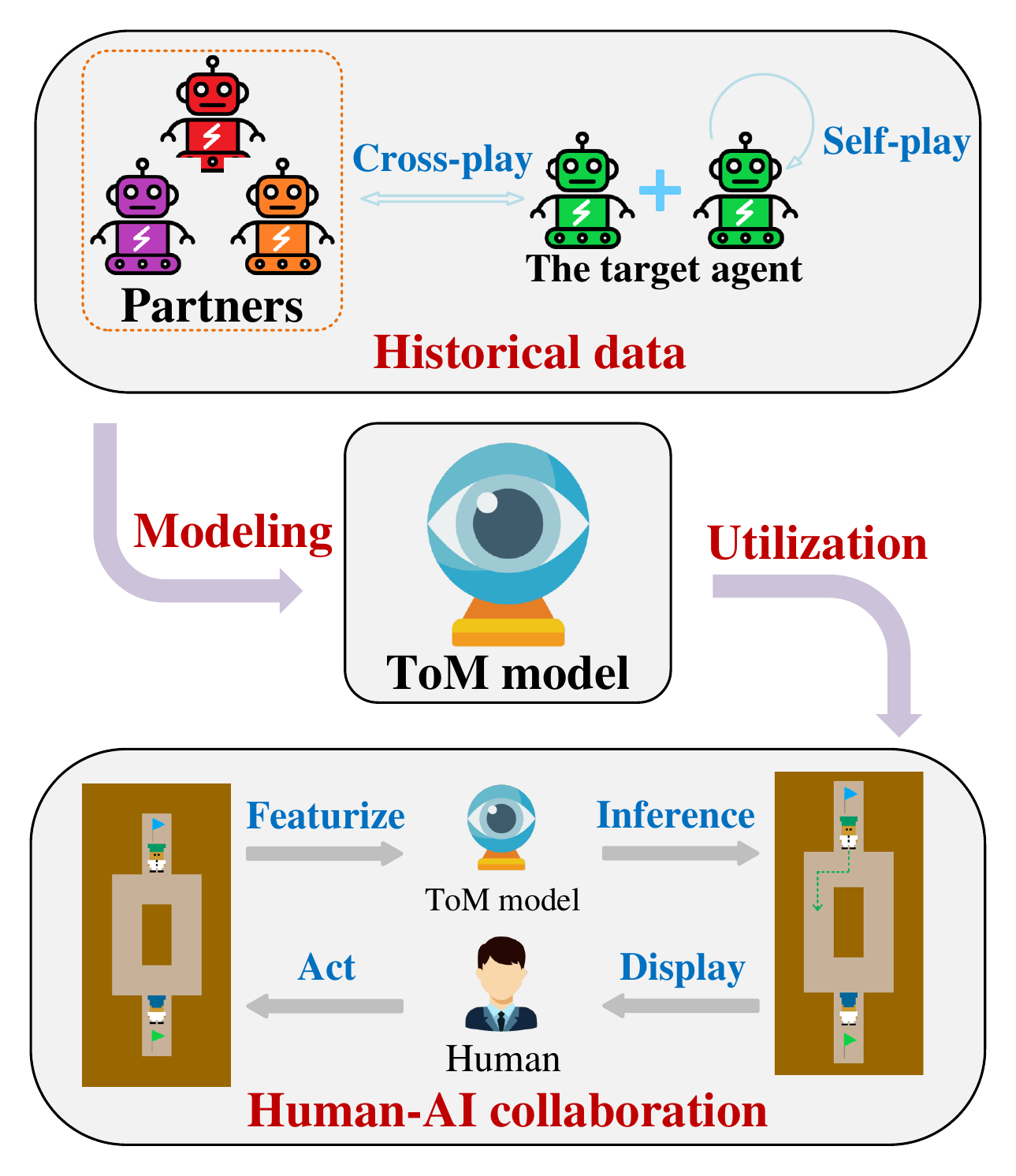}
\caption{An illustration of our two-stage paradigm for human-AI collaboration.}
\label{fig: architecture}
\end{figure}

Carrying on this idea, we propose a two-stage paradigm to assist humans in the human-AI collaboration as illustrated in \Cref{fig: architecture}. For the agent we want to predict, which is denoted as the target agent hereinafter, we first collect the trajectories by pairing the target agent with other agents to form a dataset, and then train a stand-alone ToM model from its historical data to model the behavior of the target agent. In the human-AI collaboration stage, we utilize the ToM model to predict the future action sequence of the target agent from the states of the ongoing task. The predictions are visually displayed in real-time to help the human better understand the intention of the AI agent, which can further benefit the efficiency of the collaboration.

Our proposed paradigm is distinguished from existing techniques such as plan communication \citep{porteous2023communicating, maccio2022mixed} by the following characteristics: (1) our paradigm does not require any prior knowledge of the environment, the prediction is at the action level, ensuring its availability in general DRL scenarios. (2) The ToM model is trained from offline data and can be regarded as a complete post-process that has no effect on the behavior of the target agent, providing compatibility for all DRL algorithms. (3) In our paradigm, the agent can be regarded as a gray box or even a black box, which establishes the potential of the ToM model to be a third-party assistant for practical applications.

We assume the ToM model can assist humans to be better aware of the intentions of their AI partners and coordinate better. To evaluate our paradigm and test our hypotheses, we first implement a transformer-based model to predict the desired future action sequence. We then develop an extended online experimental platform that is capable of displaying the prediction and user study based on the Overcooked environment \cite{bcp}, which is a two-player common-payoff collaborative game that has been widely used for studying human-AI collaboration and zero-shot collaboration \cite{fcp, MEP, charakorn2020investigating, yang2022optimal}. Our paradigm and model are comprehensively tested with two DRL methods in extensive human-AI collaboration experiments across multiple layouts. The results demonstrate that our work can help humans better grasp the intentions of AI partners to enhance their situational awareness and ultimately improve collaboration performance in various situations.

We summarize our main contributions as follows: 
\begin{itemize}
    \item We propose a novel human-AI collaboration paradigm that incorporates an external ToM model to predict the intention of AI agents for humans to facilitate the coordination of human-AI teams.  
    \item We implement a transformer-based predictor and design a framework to train such a model. The predictor is capable of predicting the sequence of future actions without access to the target agent.
    \item We extend the Overcooked environment to support the visual presentation of intentions and set up an online experiment platform for human-AI collaboration experiments. 

    \item We perform extensive experiments to validate our method with two types of DRL agents. The results demonstrate the utility of adopting an external ToM model in human-AI collaboration and our analysis reveals the underlying mechanism of the improvements. 

\end{itemize}

This paper is organized as follows: in \Cref{section: related work}, we review recent works that are related to our study. The implementation of our methods, including problem formulation, data collection process, and the destination of the ToM model are introduced in \Cref{section: Methodology}. In \Cref{section: Experiments}, we expatiate the experiment designation and details of the environment. We then present quantitative results and subjective user assessment with corresponding analysis. Finally, we make a conclusion and discuss the limitations and future works in \Cref{section: Conclusions}. 

\section{Related Work}
\label{section: related work}
\subsection{Human-AI collaboration}
As a frontier of AI application, human-AI collaboration has recently attracted much attention from researchers. In practice, Deep Reinforcement Learning has been proven to be a decent solution for training AI agents, which empirically shows better performance \cite{siu2021evaluation, bcp}.
Though self-play may train agents with good capability, they do not perform well in the cooperative setting. To address this problem, many studies concentrate on mitigating the disparity between training data and real collaboration by finding appropriate partners to train with. \cite{bcp} shows that training with behavior cloning models of humans as partners can achieve better performance. Without incorporating human data, treating the human-AI collaboration task as a special case of zero-shot collaboration and improving the general cooperation ability of the agent is also a popular and promising direction. Several prior works train the agent with a diversified group of pretrained partners \cite{fcp, MEP, charakorn2020investigating, lupu2021trajectory, lou2023pecan}, urge the agent to accommodate various strategies \cite{lucas2022any}, or utilize the underlying symmetries \cite{OP}.

Another direction of improvement is modeling the teammates to infer and make use of their intentions \cite{qi2018intent, choudhury2019utility}, which has fruitful achievements in the human-robot collaboration \cite{chen2021visual, luo2019human, wang2022co}. Some recent studies employ the Theory of Mind (ToM) to update the belief of other agents' intents through Bayes inference and accordingly adjust their plans \cite{lim2020improving, wu2021toomanycooks}. Beyond Bayes, ToM models can also be implemented by neural networks, which show their advantages in many multi-agent collaboration scenarios \cite{rabinowitz2018machine, wang2021tom2c}. 

However, most of the studies focus on improving the capability of AI or helping AI to better model and utilize the intents of teammates, few of them care about assisting humans to understand the intention of AI for better collaboration.

\subsection{Communication in Human-AI teaming}
Communication, as the bond that connects multiple individuals, has been widely recognized as important for Human-AI Collaboration and studied in various aspects \cite{ajoudani2018progress, yuan2022situ, ferrari2022bidirectional}. \cite{tabrez2022descriptive} shows that visual communication can effectively improve the shared situational awareness in human-robot teaming.
From the content aspect, delivering the intention of AI to humans is also known as helpful for collaboration. \cite{maccio2022mixed} shows that unilateral communication that depicts an image of the robot's future plan through a head-mount display could significantly enhance collaboration performance.
\cite{paleja2021utility} studies the utility of explaining the decision-making logic of AI and finds that explanation can improve situational awareness and further help the human AI team to achieve better performance. \cite{porteous2023communicating} shows the effectiveness of visualizing the branched plans for human-agent decision making.

However, most communication is on the basis of the explainability of the agent since the information must be conveyed in human-compatible manners, \eg, in visual or textual form. Therefore, the existing works either require rule-based models which are relatively simple \cite{maccio2022mixed, rosen2020communicating, newbury2022visualizing}, or require specified algorithms \cite{yuan2022situ, edmonds2019tale, paleja2021utility}, or rely on the communication mechanism embed in the environment \cite{gaotowards}. 
Generally, the powerful DRL methods are not universally transparent and lack explainability \cite{heuillet2021explainability}, therefore are not directly applicable to communication.

Our communication form which visually displays future actions has been widely adopted and verified by previous works. The contribution of our proposed paradigm lies in filling the gap of communicating the intentions of black-box and unexplainable agents. 

\subsection{Explainable AI}
To help humans better understand AI, explainable AI (XAI) has been widely studied \cite{gunning2019xai, xu2019explainable}, and is also extensively utilized in human-AI teaming \cite{tabrez2019improving, tabrez2019explanation, ibrahim2023explanations}.
A classical approach is to adopt policies with intrinsic explainability, such as using a decision tree \cite{paleja2021utility} or planning-based methods \cite{yuan2022situ}. Another schema is to generate a human-comprehensible explanation of the motivation and decision-making factors through specifically design models, \eg, generating a caption for autonomous driving \cite{jin2023adapt} or explaining the reward of RL-based policies \cite{iucci2021explainable}. 
However, most existing works focus on explaining the decision-making motivation instead of the future intention. Though planning-based or symbolic algorithms may have explainable behaviors, they may not perform as well as those based on deep learning.

Generally, one can not expect the AI agent to have explainable intentions, especially in complex scenarios and with DRL-based agents. In contrast, the external intention predictor in our paradigm has no limitation on the algorithm of the agent, which can be regarded as a universally available plug-in for human-AI collaboration.

\begin{figure*}[t]
\centering
\includegraphics[width=0.9\textwidth]{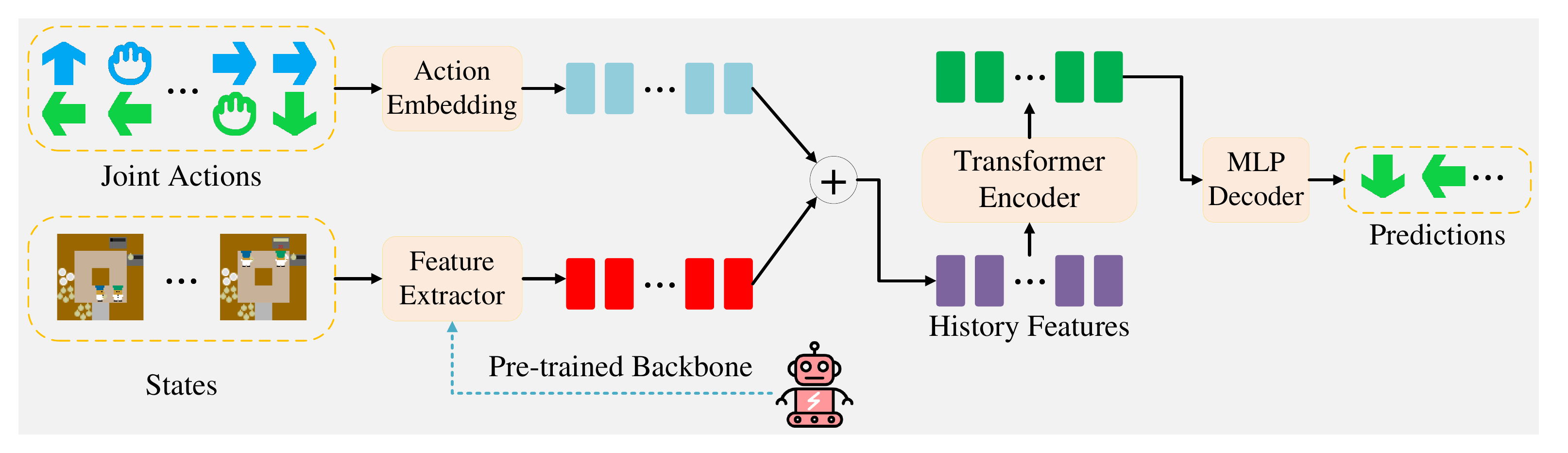}
\caption{An overview of the architecture of the prediction model.}
\label{fig: model}
\end{figure*}

\section{Methodology}
\label{section: Methodology}
In this section, we introduce the implementation details of the ToM model in our paradigm shown in \Cref{fig: architecture}, which are trained from the collected offline data and predict the future actions of the agent from the trajectory of the ongoing game at the inference stage.

\subsection{Preliminaries} \label{subsection: Preliminaries}
A typical two-player Markov Decision Process (MDP) is defined as a tuple $<S, \{ \mathcal{A}^{i}\}, \mathcal{T}, R>$, where $S$ is a finite set of states. $\mathcal{A}^{i}$ is the finite set of actions available to agent $i$, where $i \in \{1, 2\}$. The joint action space is written as $\mathcal{A} = \mathcal{A}^1 \times \mathcal{A}^2$, and the joint action is denoted as $a = (a^1, a^2)$,  where $a^1 \in \mathcal{A}^1$ and $a^2 \in \mathcal{A}^2$ are the actions of the two agents. $R$ is a real-valued reward function, and $\mathcal{T}: S \times \mathcal{A} \mapsto S $ is the transition function that maps the current state and the joint actions to the next state. 
In collaboration settings, both agents share the same reward and are trained to maximize the expected reward $\mathbb{E}_{\tau} \left[ \sum_{t}^{T} R(s_t, a_t) \right]$, where $\tau$ is the trajectory, $T$ is the number of total time steps, and $s_t$ and $a_t$ are the state and the joint action at $t$-th time step, respectively. 
For human-AI collaboration, one of the agents will be controlled by humans, and the other will be controlled by a trained policy $\pi$. The action space of the human and the AI are denoted as $\mathcal{A}^{H}$ and $\mathcal{A}^{AI}$, respectively.

\subsection{Problem Setting} \label{subsection: problem_setting}
Suppose that we have a trained policy $\pi$ which will cooperate with humans later and is denoted as the target agent. We aim to train a ToM model $F$ with trainable parameters $\theta$ from the offline trajectories to predict the future actions of $\pi$. The model $F$ is expected to take the trajectory of the current game as input and outputs a sequence of action predictions. The input trajectory can be formed as $(H, s_{n})$, where $n$ is the length of trajectory, $H = \{ (s_1, a_1), \dots, (s_{n-1}, a_{n-1}) \}$ is the history observation-action pairs and $s_{n}$ is the current observation. 
The prediction sequence is denoted by $\bm{y} = \{y_1, \dots, y_l\}$, where $l$ is the expected prediction length, and $y_l \in \mathcal{A}^{AI}$ is the predicted actions.
Note that the ToM model is assumed to be external and has no access to the action that $\pi_{A}$ will take for the current step, thus the prediction starts from the current step, \ie, $y_1$ corresponds to $a_n$.

We train the ToM model via supervised learning, where an offline trajectory dataset $D$ is built for training. The task can then be formulated into an optimization problem to maximize the log-likelihood:
\begin{align}
    \max_{\theta} {\rm{~~~~}} \mathbb{E}_{(H, s_{n}, y) \sim D} {\rm{~~log}} F_{\theta}(y | H, s_{n}).
\end{align}
However, such a problem may be difficult to optimize due to the sparsity of the label space, which grows exponentially with the length of predictions. Therefore, we simplify the problem to the action level by ignoring the temporal correlations among actions and to get an approximate solution as a baseline. The reduced problem is written as:
\begin{align}
    \max_{\theta} \sum_{(H, s_{n}, y) \in D} \sum_{i=1}^{l}  {\rm{~~log}} F_{\theta}^{(i)}(y_i | H, s_{n}),
\end{align}
where $F_{\theta}^{(i)}$ denotes the model for the $i$-th prediction. We further implement a model as an approach to this problem.

\subsection{Data Collection}
As high-quality data is vitally important in deep learning, building a diversified and representative dataset is the first step in training the ToM model. However, collecting human data is too expensive to be a practical solution in both temporal and economical sense. Inspired by fictitious co-play (FCP) \citep{fcp}, we find that deliberately chosen AI partners can also provide diversified data to improve the generalization performance of the model and further benefit the human-AI collaboration. 

Our data collection process is presented in \Cref{algo: collect_data}, where we gather data from both self-play and cross-play. In self-play, the target agent plays with a copy of itself, where the trajectories demonstrate the ideal plan in the agent's opinion. In cross-play, the target agent is paired with a group of partners to obtain the behavior pattern when cooperating with others, where collaboration may not be ideal. Specifically, we train an independent group of agents with various DRL algorithms and pick 3 checkpoints of each agent with different skill levels to enrich the diversity in the skill sense and form the partner population. Different from FCP \citep{fcp}, we deem humans to have at least a basic level of skill and choose the checkpoints that achieve 35\%, 70\%, and 100\% rewards of the best checkpoint. 
For each selected pair of agents, we collect trajectories under different settings, including the roles of agents and whether the partner agent is deterministic. 


\begin{algorithm}[t]
\caption{Data collection process} \label{algo: collect_data}

\begin{algorithmic}[1]
\Statex \textbf{Input: The MDP $M$, target policy $\pi$, game setting set $\mathcal{C} $}
\Statex \textbf{Output: Trajectory dataset $D$}

\State $D \gets \emptyset$
\State $\mathcal{P} \gets \{\pi\}$
\State $agents \gets $ train\_agents($M$)

\For {\rm{\textbf{each}} $agent \in agents $}
\State $checkpoints \gets$ select\_checkpoints(agent)
\State $\mathcal{P} \gets \mathcal{P} \bigcup checkpoints$
\EndFor

\For {\rm{\textbf{each}} ${\pi}_{p} \in \mathcal{P} $}
\For {\rm{\textbf{each}} $c \in \mathcal{C} $}
\State $t\gets$ get\_trajectory(${\pi}_{p}$, $\pi$, $c$)
\State $D \gets D \bigcup t$
\EndFor
\EndFor

\Statex \textbf{Return: $D$}
\end{algorithmic}
\end{algorithm}

\subsection{Model Implementation}
We present the architecture of our ToM model in \Cref{fig: model}.
To utilize deep learning models, the past trajectory $(H, s_{n})$ is first decoupled into a sequence of joint actions and a sequence of states, denoted as $\{a_1, \dots, a_n\})$ and $\{s_1, \dots, s_n\})$, respectively. Note that the action at the current step is unknown, thus $a_n$ is padded as a zero-vector. 
The actions of both characters are first converted into the one-hot representations, then concatenated into a vector with a dimension of $|\mathcal{A}|$. We encode the actions through a linear layer to get the embeddings. Similarly, the states are encoded by a feature extractor, which is a Convolutional Neural Network (CNN) in practice. Specifically, we train a separate agent and utilize its feature extractor as the pre-trained backbone, which offers a decent start of training while keeping the black box assumption of the target agent. 

The action embeddings and state features are then concatenated to get a unified representation of history features, denoted as $H' = \{h_1, \dots, h_n\}$. Next, we employ a transformer model to encode the history feature sequence $H'$ and decode the hidden representation to get the prediction $\hat{y}$ via a decoder. Since our application scenario only needs fixed-length predictions, we use an MLP as the decoder instead of auto-regressive models. As described in \Cref{subsection: problem_setting}, each element of the prediction sequence is inferred independently, so the Softmax function is used as the activation function of the last layer of the MLP decoder.
Finally, the model is optimized by minimizing Cross-Entropy loss, which has the following form: 
\begin{align}
    \mathcal{L} = -\sum_{i=1}^{l} \sum_{j=1}^{|\mathcal{A}^{AI}|} y_{ij} log(\hat{y}_{ij}).
\end{align}
Where $\hat{y}_{ij}$ denotes the probability of $j$-th action, $y_{ij}$ is 1 for the correct action and 0 otherwise.

\section{Experiments}
\label{section: Experiments}

\subsection{Environment}
To inspect the effectiveness of our human-AI collaboration framework, we conduct experiments in the Overcooked environment \cite{bcp}, which has been widely used for studying zero-shot coordination and human-AI collaboration \cite{bcp, charakorn2020investigating, knott2021evaluating, fcp, MEP, yang2022optimal, lou2023pecan}.
In Overcooked, players should coordinate to pick up ingredients, put the ingredients into pots to cook, dish out the soup when it is done, and serve it to the designated location. In our experiment, the pot automatically starts cooking when it has 3 onions and it takes 20 time steps to cook. All the players get the same reward when a dish is successfully served, thus the mutual goal is to serve as many soups as possible in a limited time. 
There are 6 available actions for each player: $move$ $\{up, down, left, right\}$, $interact$, and $wait$. A player can not move to the cell where the other player is already there. If the two players try to go to the same grid at the same time, both of them will fail to move. To perform well, the team needs to coordinate both high-level strategies and low-level path planning. 

\begin{figure}[t]
\centering
\includegraphics[width=0.47\textwidth]{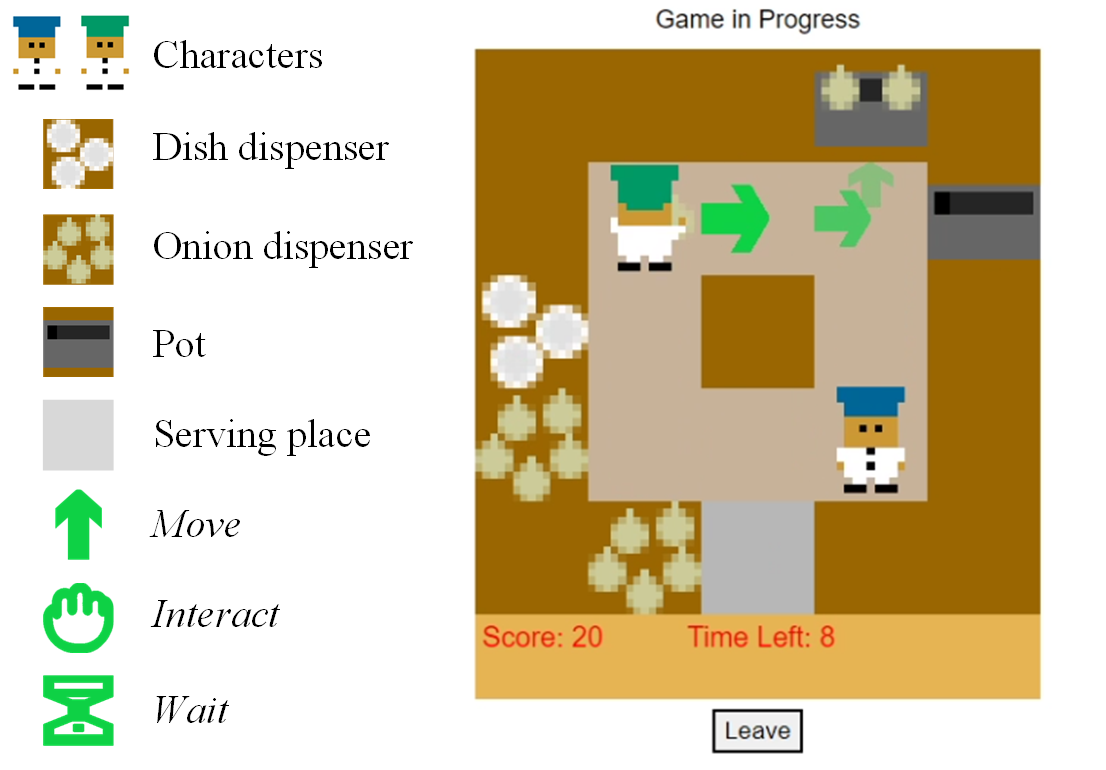}
\caption{Our extended experiment platform based on Overcooked \citep{bcp}. We present the illustration of the icons and a screenshot of the user interface with shown predictions.}
\label{fig: environment}
\end{figure}

To provide action predictions in the human-AI collaboration process, we extend the front-end user interface to show predictions by depicting successive arrows and icons with gradually varied sizes and transparency at estimated locations. Our extended version of Overcooked is illustrated in \Cref{fig: environment}. To avoid confusion, we always let the human control the blue character and let the AI control the green one. In the given example, the arrows correspond to the prediction sequence $\{move\ right$, $move\ right$, $move\ up \}$, which may imply the AI is going to put its holding ingredient into the top-left pot.

\begin{figure*}[t]
\centering
\includegraphics[scale=0.14]{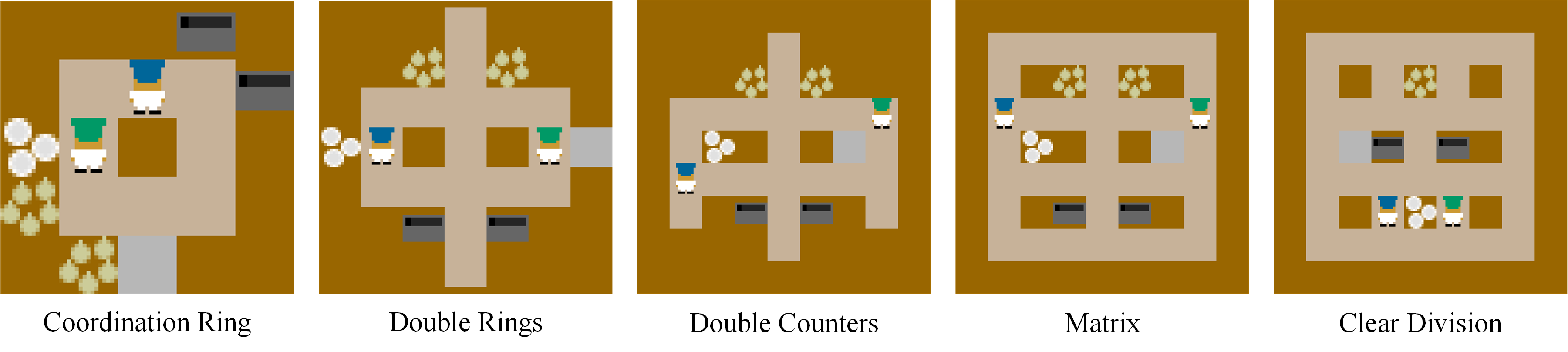}
\caption{The adopted Layouts. We use 5 various layouts to test our paradigm in different circumstances. \textit{Coordination Ring} challenges the ability of the human-AI team to cooperate in a narrow space. In \textit{Double Rings}, the team must coordinate in both path choices and task allocation. \textit{Double Counters} is a variant of \textit{Double Rings} that reduces the difficulty of coordination while increasing the punishment of conflicts. \textit{Matrix} is designed as a complicated environment where exists multiple reasonable coordination solutions. Finally, \textit{Clear Division} has an intuitive plan for task allocation and low difficulty in operation though it is large.}
\label{fig: layouts}
\end{figure*}

We use 5 different layouts as illustrated in \Cref{fig: layouts} in the experiment, which are expected to offer various challenges. 



\subsection{Experimental setup}
To comprehensively study the utility of our ToM model, we incorporate two types of agents in the experiment: Self-play (SP) and Fictitious Co-Play (FCP) \cite{fcp}. Following the results of prior works, the SP agent is regarded as strong but egocentric, while FCP is more cooperative and performs better in zero-shot collaboration scenarios. We expect the ToM model can help in both situations.

For the agents, we use the same designation as in \cite{bcp} which has a 3-layer CNN as the feature extractor (the kernel sizes are $5 \times 5$, $3 \times 3$, and $3 \times 3$, respectively) and a 3-layer MLP to output the actions. We use the PPO algorithm for the optimization with hyper-parameters that are generally consistent with \cite{bcp}. Each agent is trained for $1e7$ time steps with a learning rate of $1e-3$. More training details can be found in our provided supplementary. For both SP and FCP, we train 3 agents and select the one with the best performance for the human-AI collaboration experiment. 

To train the ToM models, we first train 5 partner agents in both SP and FCP and randomly divide them for training, validation, and testing, with a proportion of 3:1:1. Then we pair the target agent with each partner to collect the trajectory dataset. We collect 5 trajectories of 800 steps for each pair of agents following the procedure described in \Cref{algo: collect_data}, resulting in $4.96e5$ steps of data.
We train an individual FCP agent and take its trained CNN as the pretrained feature extractor for all ToM models. In our experiment, the length of history is $10$ and the prediction length is set to $3$. The transformer is encoder-only, having 4 stacked transformer layers, each with 8 attention heads, and the hidden size is set to $128$. The predictor is trained for 100 epochs under a learning rate of $1e-3$ with an early stopping on the validation set.

\begin{figure*}[t]
\centering
\includegraphics[width=\textwidth]{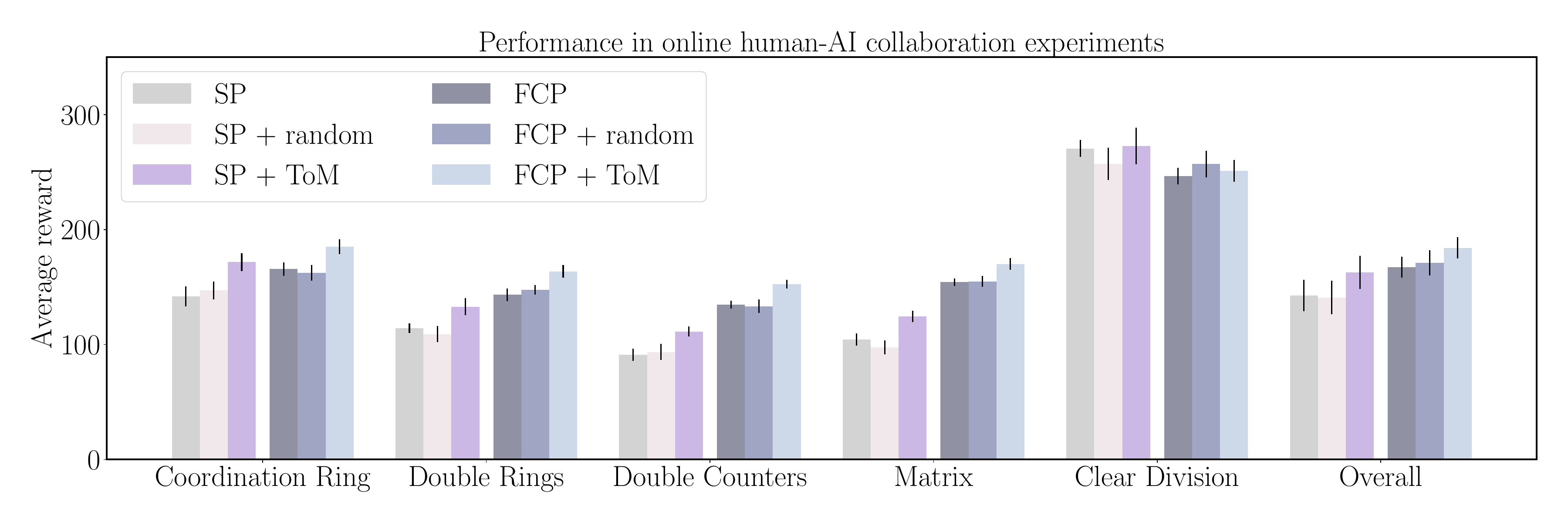}
\caption{Average rewards with standard error bar. Each successful delivery gains 20 to the reward. To make the experiment more fine-grained, we give a discount reward according to the process of making soups at the end of the game. To reduce randomness, the initial positions are fixed in each layout.}
\label{fig: scores}
\end{figure*}

\subsection{User study designation}
We run an online human-AI collaboration experiment to test our paradigm and models. Our experiment goes through an internal ethics review. All participants are informed of the experiment information, including the process, compensation, and privacy protection. Explicit informed consent is required for participation. We recruited 87 participants (42 female, 45 male, ages 18-55) in various ways for the experiment. 

Since our main purpose is to study the utility of predictions, we introduce two baselines as the control group: no prediction and random prediction. We use a between-subject designation, in which the participants are divided into 3 groups, playing with one setting of predictors respectively. Each participant only plays with the same class of predictor throughout the experiment. Participants first read an introduction to the Overcooked game and are required to pass a tutorial game. They then play with the two agents (SP and FCP) in the 5 layouts, in total 10 episodes. The order of layouts is the same as illustrated in \Cref{fig: layouts}. The order of agents in the first layout is randomly assigned, and the others are reversed with respect to the predecessor. Each episode lasts 400 time steps (80 seconds). Compared to previous work \cite{bcp}, we set a lower game speed (5 fps) to give the participants more room to play. 
After finishing each episode, participants are asked to give a subjective assessment by filling out a questionnaire with the following questions:
\begin{itemize}
    \item How satisfied are you with your AI teammate? 
    \item How satisfied are you with the predictor (if applicable)?
    \item To what extent can you predict the intention of your AI teammate?
    \item How would you rate the efficiency of cooperation? 
\end{itemize}
The four questions correspond to \textit{satisfaction with the partner}, \textit{satisfaction with the predictor}, \textit{situational awareness}, and \textit{cooperation efficiency}, respectively. 
Participants are asked to fill in an integer between 1 and 7 for each question.

An online platform is developed for the experiment, which provides a set of interfaces for the administrator to monitor the experiment. After the collaboration, we check the trajectories and remove clearly broken games (about $1.1\%$). Finally, we got 860 episodes of data for analysis.

\begin{table*}[th]
    \centering
    \caption{Prediction accuracy (\%) of the ToM model}
    \begin{tabular}{ccccccc}
        \hline
        Agent & Dataset & Coord. Ring  & Double Rings & Double Counters & Matrix & Clear Division  \\
        \midrule
        \multirow{2}{*}{SP} & Offline test & 79.42 & 70.91 & 73.70 & 76.08 & 73.44 \\
        \cline{2-7} 
        & User study & 82.29 & 61.77 & 65.84 & 60.37 & 77.88 \\
        \hline
        \multirow{2}{*}{FCP} & Offline test & 89.54 & 87.61 & 85.40 & 76.83 & 91.44 \\
        \cline{2-7} 
        & User study & 86.60 & 78.34 & 76.51 & 76.97 & 83.13 \\
        \hline
        
    \end{tabular}
    \label{tab: pred_acc}
\end{table*}

\subsection{Results}
\subsubsection{Quantitative performance} \label{sec: results}
The most immediate indicator is the reward in each episode, thus we present the average reward on each layout and across layouts in \Cref{fig: scores}. We first perform Kruskal-Wallis H tests (a non-parametric one-way ANOVA) to compare the rewards in different ToM settings for both SP and FCP agents. The results show statistically significant differences in rewards for both SP ($H(2, n=430)=10.83, p=.004$) and FCP ($H(2, n=430)=15.23, p<.001$). 
Subsequent ANOVA conducted on the results in each layout show that significant differences exist on the first 4 layouts (with a significance level of $p < .05$).
Since we are mainly curious about the efficacy of our trained ToM model, we run pairwise comparisons between the reward acquired in each layout with each agent with or without the ToM model with the T-test. Even after the Benjamini-Hochberg correction for false discovery control, the ToM model still significantly improves the rewards in the first 4 layouts (within the significance level of $p < .05$).

The results corroborate that our ToM model can benefit human-AI collaboration in various situations. Such improvements cover multiple layouts, which implies the predictions help the collaboration in both high-level strategy coordination and low-level routine coordination. Furthermore, the ToM model is helpful for both the SP agent and FCP agent, though the FCP agent has already significantly outperformed the SP agent and achieved a relatively high reward. We deem that the ToM model helps the collaboration in another aspect, which is orthogonal to the ability of agents. Besides, the random baseline confirms that the improvement is not caused by any placebo effect, and the visual presentation itself has no remarkable impact on the human-AI collaboration. 

We are also interested in the negative results, where the ToM model fails to improve the performance on \textit{Clear Division}. The performance in all three settings has no significant difference from the others. Qualitatively, \textit{Clear Division} has a straightforward scheme for high-level task allocation: one player goes top to put ingredients into the pots, and the other is responsible for serving cooked soup. In such a task allocation schema, the routes of the two players have almost no intersections, which further eliminates the challenge of low-level motion coordination. We subjectively analyze the playback of the experiment and find that most teams adhere to such a scheme, thus no room for improvement is left for the ToM model.
This perspective is further substantiated by the comparison between SP and FCP agent performances, which demonstrated no notable advantage for the latter though it is more cooperative.

\begin{figure}
    \centering
    \includegraphics[width=0.47\textwidth]{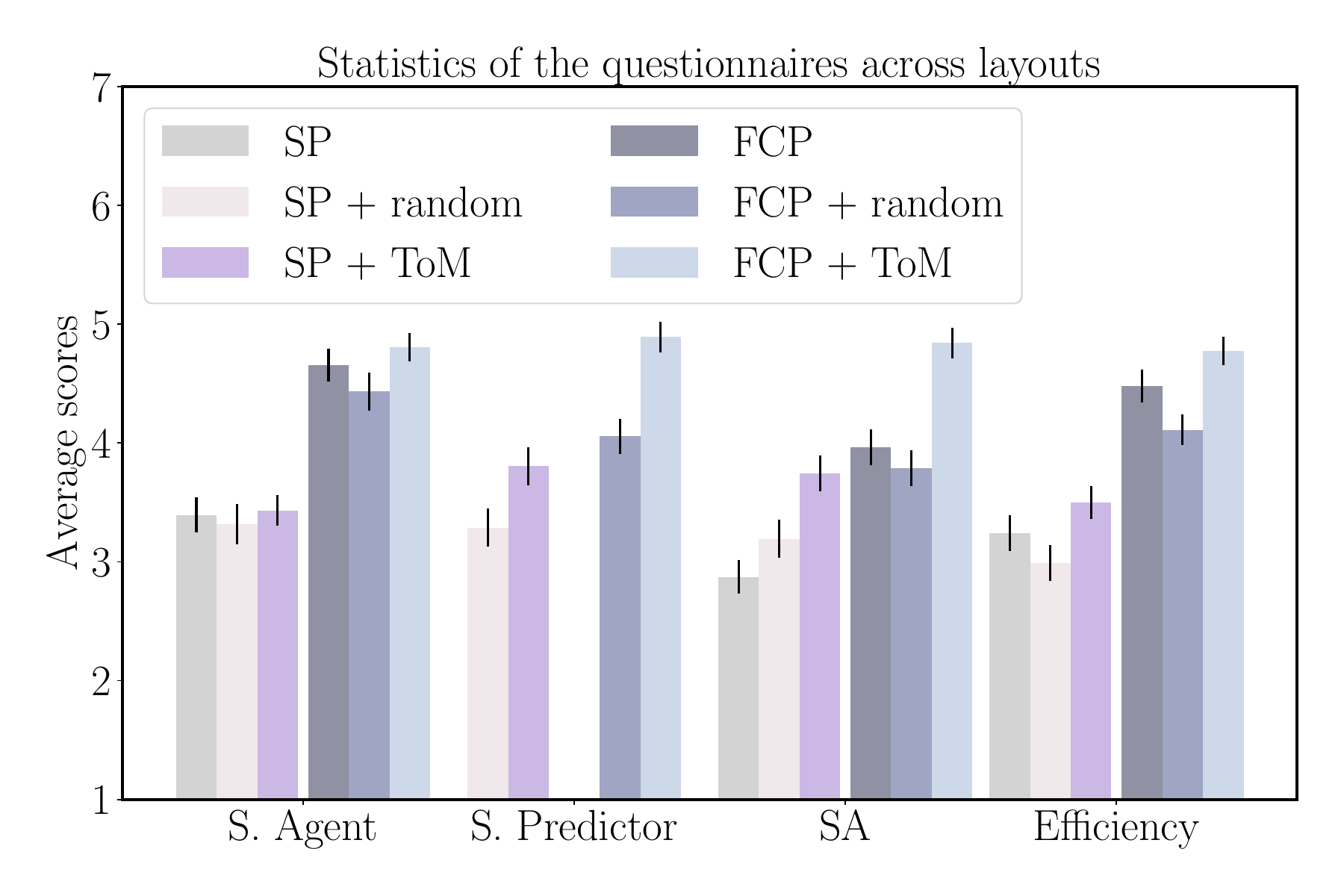}
\caption{Average scores in user assessment across layouts.}
\label{fig: questionnaires}
\end{figure}

\subsubsection{User assessment}
To test the hypothesis that the improvement in coordination is the main factor of performance improvements, we scrutinize user assessments and present the average scores across settings in the \Cref{fig: questionnaires}. Since we aim to investigate the factor in positive situations, only settings with significant improvement are taken into account, \ie, data on the first 4 layouts. 

As expected, the satisfaction of agents has no significant difference, because the agents are exactly the same. As for the satisfaction of the predictor, the ToM model gets significantly higher scores than the random baseline (SP: $t(224)=2.33, p=.021$; FCP: $t(226)=4.16, p<.001$), underscoring that humans care about the predictions and can identify the capability of predictors. 
The ToM model also significantly improves situational awareness compared to both the random baseline and the no-predictor setting (all with a significance level of $p<.05$), whereas the random baseline shows no significant difference from no predictor, confirming the utility of appropriate predictions.

Although we have quantitatively proved that the ToM model can improve the efficiency of human-AI collaboration, it is not fully reflected in the subjective feelings of participants. However, the difference between the ToM model and the random baseline turns out to be significant (SP: $t(224)=2.44, p=.016$; FCP: $t(226)=3.68, p<.001$), which may indicate that inappropriate predictions make humans confuse and spoil their subjective feelings.

The between-subject designation may reduce the statistical significance of the user assessment since the user can not compare among different settings of predictors and tends to give a middle score. Though the participants who cooperate with a random predictor can not forecast their partner well, they tend to give a score of 3 to 4 for \textit{satisfaction of the predictor}, since they are not told that it is a random predictor. Accordingly, this suggests that the ToM model's impact could be more substantive than the questionnaire results indicate.

\subsubsection{Prediction Accuracy}
As discussed above, performance improvement highly depends on the accuracy of predictions, random predictor leads to no improvement and even negative effect. Here we show the accuracy of our ToM model on each layout in \Cref{tab: pred_acc}, where the accuracy is calculated at the action level. The offline test refers to the test set of the collected dataset, in which the target agent is paired with DRL partners. We also test the exact accuracy of the ToM model in human-AI collaborations, denoted as user study.

Note that the future actions of the target agent also depend on its partner, \eg, the agent may fail to execute its plan if the human blocks its way. Although our ToM model can recognize if the agent is stuck from the current game history and adjust the prediction, there still exists an upper limit of accuracy which is lower than $100\%$ when the partner is non-deterministic. 
Overall, the ToM model has a decent accuracy that can well support human-AI collaboration tasks. However, compared to the offline test where collaborating with a DRL agent, there is a general decrease in accuracy in the real human-AI collaboration (8 of 10 scenarios). Such a result may indicate that humans are generally more diversified, leading to novel situations that are out of the distribution of the training data, and humans also act more randomly. This indicates further room for improvement in the designation of the predictor and the presentation of predictions. 

\subsection{Case Study}
\begin{figure}
    \centering
    \includegraphics[width=0.4\textwidth]{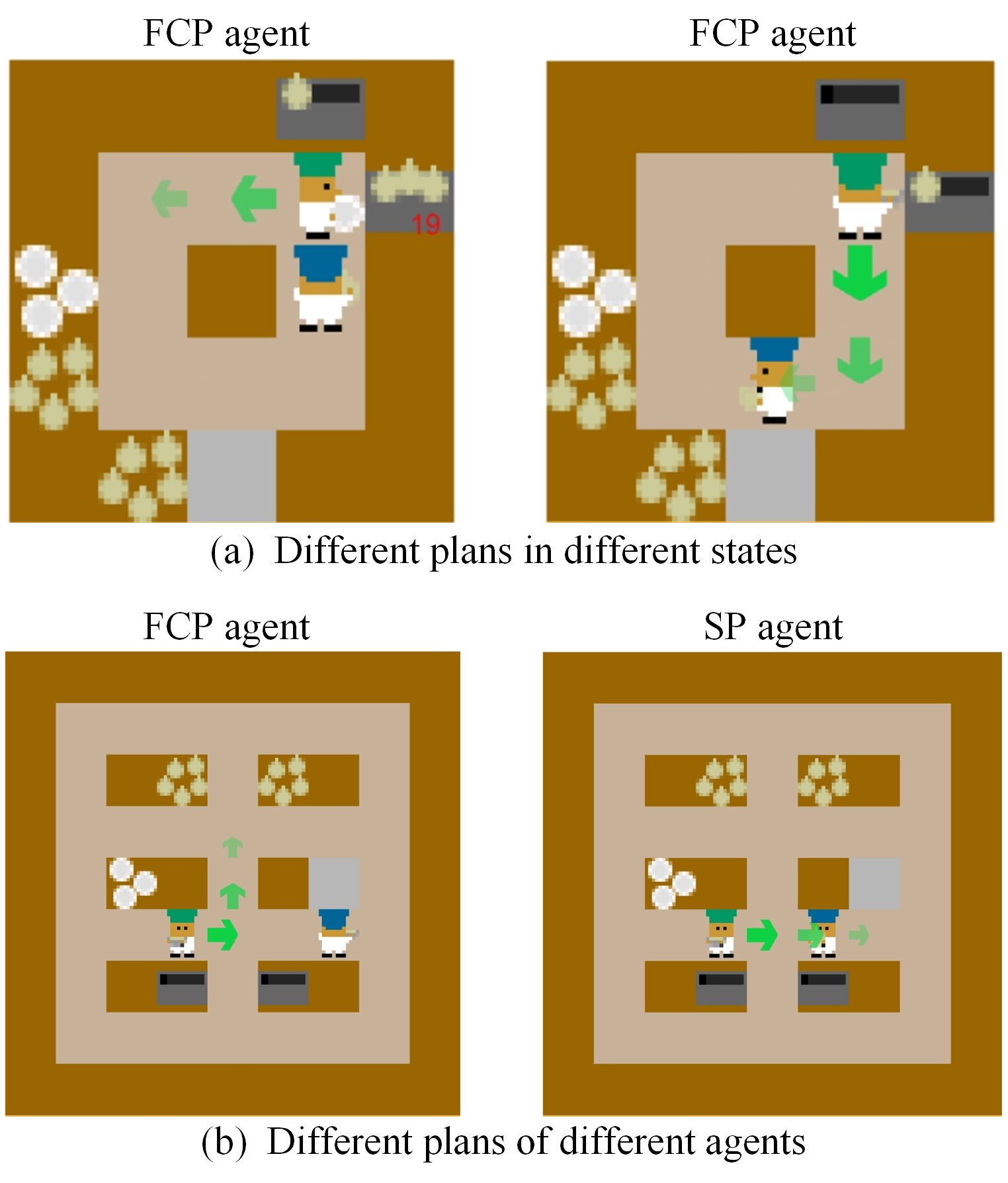}
\caption{Representative screenshots in the human-AI collaboration. In all presented cases, the future actions of the agents are perfectly predicted and indicate the plans.}
\label{fig: case_study}
\end{figure}

To qualitatively analyze the effect of the ToM model and our paradigm, we review some records of the experiment and present some representative cases in \Cref{fig: case_study} as a case study.

Naturally, the ToM model contributes the most when there exist multiple reasonable plans. In both examples in \Cref{fig: case_study}, the agent is going to the serving place and there are two reasonable ways. With the shown future actions, the human can understand the intention of the AI agent and manage to not block its way.
We argue that the ToM model is important to human-AI collaboration because the intentions of AI are hard to infer by humans. The plan of the same AI agent can vary with the situation. For instance, in \Cref{fig: case_study} (a), the shortest path towards the destination is moving downward, however, the agent chooses the left-hand way when the human is in the way, to avoid possible collisions. Moreover, even in similar cases, different agents may make different choices. As shown in \Cref{fig: case_study} (b), the FCP agent will choose the upper route whereas the SP agent chooses the lower one, disregarding that the human is in its way. Generally, the behaviors of agents are hardly predictable for humans, especially when humans are not familiar with their partners. Therefore, our paradigm which introduces an external intention predictor can help a lot in the human-AI collaboration process.

\section{Discussion}
\label{section: Conclusions}
\subsection{Conclusion}
In this paper, we introduce a new human-AI collaboration paradigm that incorporates a ToM model that trained from the offline trajectories of the target AI agent and utilized in the collaboration process. We further implement a framework in which trajectories from both cross-play and self-play to build a diversified replay dataset, and train a transformer-based ToM model as the predictor. 
We then develop an online experiment platform based on the extended Overcooked environment and design a between-subject experiment with post-game user assessment to testify our paradigm and implemented ToM models. 
The results of the extensive experiment demonstrated the helpfulness of our paradigm to human-AI collaboration in both quantitative rewards and subjective measurement. Our study of such a paradigm and implementation of the platform build a foundation for further research on assisting humans to understand the intention of agents in human-AI collaboration.

\subsection{Limitation and future work}
Our paradigm mainly contributes to the human-AI collaboration task, where the experiment results are highly related to the human subjects. Even the quantitative results highly depend on the game skills of participants. Since the participants are primarily recruited from the Internet, we have no commitment to the representativeness of the participants. Though the experiment has validated the general effectiveness of our paradigm, it may require further analysis when applied to specified realistic scenarios.

For future work, a possible direction is to study the safety and variation of human trust on the proposed external intention predictor. As implied by the random baseline in our experiment, inappropriate predictions may reduce human trust. Especially, it may be harmful if the ToM model is maliciously manipulated to mislead humans, which may cause a serious failure.

Since our ToM model only predicts the action sequence and lets humans speculate the intention from such a prediction, a possible direction of improvement is to generate semantically comprehensible intentions, such as the current goal of the agent. Besides, it should be meaningful to verify our paradigm in more complicated and practical human-AI collaboration domains, especially the domains in which the global visual display is not applicable may bring new challenges to communication.


\bibliographystyle{IEEEtran}
\bibliography{bib.bib}

\end{document}